\documentclass[twocolumn]{aastex63}

\shorttitle{Massive First-Star Binary}
\shortauthors{Sugimura et al.}

\begin{document}

\title{The Birth of a Massive First-Star Binary}

\author{Kazuyuki Sugimura}
\affiliation{Department of Astronomy, University of Maryland, College Park, MD 20740, USA}
\affiliation{Astronomical Institute, Graduate School of Science, Tohoku University, Aoba, Sendai 980-8578, Japan}
\email{sugimura@astro.umd.edu}

\author{Tomoaki Matsumoto}
\affiliation{Faculty of Sustainability Studies, Hosei University, Fujimi, Chiyoda, Tokyo 102-8160, Japan}

\author{Takashi Hosokawa}
\affiliation{Department of Physics, Kyoto University, Sakyo, Kyoto 606-8502, Japan}

\author{Shingo Hirano}
\affiliation{Faculty of Sciences, Department of Earth and Planetary Sciences, Kyushu University, Fukuoka 812-8581, Japan}

\author{Kazuyuki Omukai}
\affiliation{Astronomical Institute, Graduate School of Science, Tohoku University, Aoba, Sendai 980-8578, Japan}



\begin{abstract}
We study the formation of massive Population III binary stars using a newly
developed radiation hydrodynamics code with the adaptive mesh refinement
and adaptive ray-tracing methods.  We follow the evolution of a typical
primordial star-forming cloud obtained from a cosmological hydrodynamics
simulation. Several protostars form as a result of disk fragmentation
and grow in mass by the gas accretion, which is finally quenched by the
radiation feedback from the protostars.  Our code enables us, for the
first time, to consider the feedback by both the ionizing and
dissociating radiation from the multiple protostars, which is essential
for self-consistently determining their final masses.  At the final step
of the simulation, we observe a very wide ($\gtrsim 10^4\,\mathrm{au}$)
binary stellar system consisting of $60$ and $70\,M_\odot$ stars. One of
the member stars also has two smaller mass ($10\,M_\odot$) companion
stars orbiting at $200$ and $800\,\mathrm{au}$, making up a mini-triplet
system.  Our results suggest that massive binary or multiple systems are
common among Population III stars.
\end{abstract}



\section{Introduction} 
\label{sec:intro}

The first generation of stars in the Universe, also known as the
Population III (Pop III) stars, are born in minihalos of
$10^{5}-10^{6}\,M_\odot$ around the redshift $z\sim20\,\text{--}\,30$ in
the framework of the standard Lambda cold dark matter ($\Lambda$CDM)
cosmology (e.g., \citealp{Abel:2002aa,Bromm:2002aa}; see also
\citealp{Glover:2013aa, Greif:2015aa} for review).  Small embryonic
protostars formed by the gravitational collapse of the natal clouds
\citep[e.g.,][]{Omukai:1998aa,Yoshida:2008aa} grow in mass by accretion
of the surrounding gas \citep[e.g.,][]{Omukai:2003ab,Tan:2004aa},
finally reaching the mass of $10\,\text{--}\,1000\,M_\odot$
\citep[e.g.,][ hereafter H16]{Susa:2014aa,Hirano:2015aa,Hosokawa:2016aa}
when the radiation feedback quenches the gas supply
\citep[e.g.,][]{Omukai:2002aa,McKee:2008aa,Hosokawa:2011aa}. If born as
binaries, such massive stars can be the progenitors of the binary black
holes (BHs) with masses $\gtrsim30\,M_\odot$ recently observed in
gravitational waves (e.g.,
\citealp{Kinugawa:2014aa,Hartwig:2016aa,Abbott:2016aa}, but also see
\citealp{Belczynski:2017aa}), as well as the X-ray binaries that affect
the thermal history of the intergalactic medium
\citep{Mirabel:2011aa,Jeon:2014ab}, which is a target of future 21cm
line observations \citep[e.g.,][]{Dewdney:2009aa}.

Recent numerical simulations have shown that the disks around Pop III
protostars are gravitationally unstable and fragment into clumps, in
which other protostars can form
\citep[e.g.,][]{Stacy:2010aa,Stacy:2012aa,Stacy:2016aa,Clark:2011ab,Greif:2011aa,Greif:2012aa,Smith:2011aa,Hirano:2017aa,Sharda:2019aa}.
Although most of these protostars merge with another protostar soon
after their formation, some of them may survive for a longer period and
may end up with binary or multiple systems
\citep[e.g.,][]{Susa:2019aa,Chon:2019aa}.  Unfortunately, however, those
simulations covered only early phases of the star formation process, and
thus the fate of the protostars remains unknown.  To reveal the nature
of resulting stellar systems, we need to follow not only the formation
of multiple protostars by disk fragmentation but also their long-term
evolution under the influence of the protostellar radiation.

In this work, we perform simulation of Pop III star formation,
self-consistently taking into account the radiation from multiple
protostars, by using a newly developed radiation hydrodynamics code,
SFUMATO-RT.  We follow the entire star formation process until the
protostellar radiation feedback terminates the gas accretion to the
newborn stellar system.  In this Letter, we describe results of a run
with a typical initial condition for the primordial star formation.
Results for various cases with further analysis, together with a
detailed description of our code, will be presented in a forthcoming
publication (Sugimura et al., in preparation).

\section{Numerical Method} 
\label{sec:methods}

\subsection{Radiation hydrodynamics simulation} 
\label{sec:code}

We use a new radiation hydrodynamics code, SFUMATO-RT, which is a
modified version of a self-gravitational magnetohydrodynamics code with
adaptive mesh refinement (AMR), SFUMATO
\citep{Matsumoto:2007aa,Matsumoto:2015ab}. We have newly added a
chemistry module coupled with radiation transfer (RT) to consider the
thermal evolution of the primordial gas under the influence of the
radiation from protostars. The hydrodynamical scheme has second-order
accuracy in space and time.

The model of primordial chemistry and thermodynamics is basically the
same as in H16. We solve the non-equilibrium chemical reactions among
six species, $\mathrm{H}^+$, $\mathrm{H}$, $\mathrm{H}_2$,
$\mathrm{H}^-$, $\mathrm{H}_2^+$, and $\mathrm{e}^{-}$, assuming all
$\mathrm{He}$ to be neutral. We consider chemical reactions and
heating/cooling processes relevant in the density range
$n_\mathrm{H}<10^{13}\,\mathrm{cm^{-3}}$, where $n_\mathrm{H}$ is the
number density of hydrogen nuclei.

Protostars are represented by sink particles, which interact with the
gas through gravity and accretion.  The threshold density for particle
creation is set at
$n_\mathrm{sink}=2\times10^{11}\,\mathrm{cm^{-3}}$. The particles
accrete the gas within the sink radius
$r_\mathrm{sink}=64\,\mathrm{au}$, which is equal to a half the Jeans
length for molecular gas with $n_\mathrm{H}=n_\mathrm{sink}$ and
$T=1000\,\mathrm{K}$.  Particles are assumed to merge when their
distance becomes shorter than $2\,r_\mathrm{sink}$.

We calculate the radiative property of Pop III protostars using a
pre-calculated table obtained with a one-dimensional stellar evolution
code under the assumption of constant accretion rates
\citep{Hosokawa:2009aa,Hosokawa:2010aa}.  The table gives the emission
rates of extreme-ultraviolet (EUV; $h\nu>13.6\,\mathrm{eV}$) ionizing
and far-ultraviolet (FUV; $11.2\,\mathrm{eV}<h\nu<13.6\,\mathrm{eV}$)
dissociating photons for a given set of the stellar mass $M$ and
accretion rate $\dot{M}$.  We average the accretion rates over 300
years, to take into account the transport of material through the
unresolved parts of the accretion disks, as in H16. In addition, no
strong time variation of $\dot{M}$ is observed in the late phase of our
run when the radiative feedback is significant partly because protostars
acquire mass through disk accretion driven by the gravitational torque
of spiral arms rather than through mergers of clumps
\citep[see][]{Stacy:2016aa}. We can therefore neglect the dependence of
stellar properties on the accretion history (but see also H16 for the
case the accretion history matters).

The RT of the direct photons from each protostar is solved with the
adaptive ray-tracing (ART) method
\citep{Abel:2002ab,Wise:2011aa,Rosen:2017aa,Kim:2017aa}, with which the
rays are adaptively split to ensure the minimum number of rays per each
cell, $N_\mathrm{ray}=3$. Along each ray, we calculate the absorption of
EUV photons by H ionization and the FUV photons by H$_2$ self-shielding,
as in H16.

Our computational domain is a cube with the side length
$L_\mathrm{box}=5\times10^5\,\mathrm{au}$. We set the base grids with
$N_\mathrm{base}=128$ cells in each direction and adaptively refine the
cells to resolve one Jeans length with at least 16 cells.  We set the
maximum refinement level $l_\mathrm{max}=10$, resulting in the minimum
cell size of $\Delta x_\mathrm{min}=L_\mathrm{box}/N_\mathrm{base}
\times 2^{-l_\mathrm{max}}\approx 4\,\mathrm{au}$.

\subsection{Initial condition}
\label{sec:IC}

We simulate the formation of a Pop III star binary/multiple system in a
fully cosmological context. To this end, we pick up a typical primordial
star-forming cloud from more than 1600 samples obtained in the previous
cosmological 3D SPH/$N$-body simulations
\citep{Hirano:2014aa,Hirano:2015aa}. The cloud we have chosen is the
same as the case C of H16 , for which they found the mass of the formed
star has the median value among the five cases examined. This cloud
begins to collapse at $z=24$ in a minihalo of $2\times10^{5}\,M_\odot$.
We start our radiation hydrodynamics simulation by remapping the
particle-based cosmological simulation data to our Cartesian grids when
the central density reaches $10^6\,\mathrm{cm^{-3}}$. We stop the
simulation at $1.2\times10^5\,\mathrm{yr}$ after the first protostar
formation, when the accretion is almost terminated and the final stellar
masses are fixed.  At the end, the simulation has required 9 months with
512 cores.

\section{Results} 
\label{sec:results}

\begin{figure}
\centering
\includegraphics[width=8.4cm]{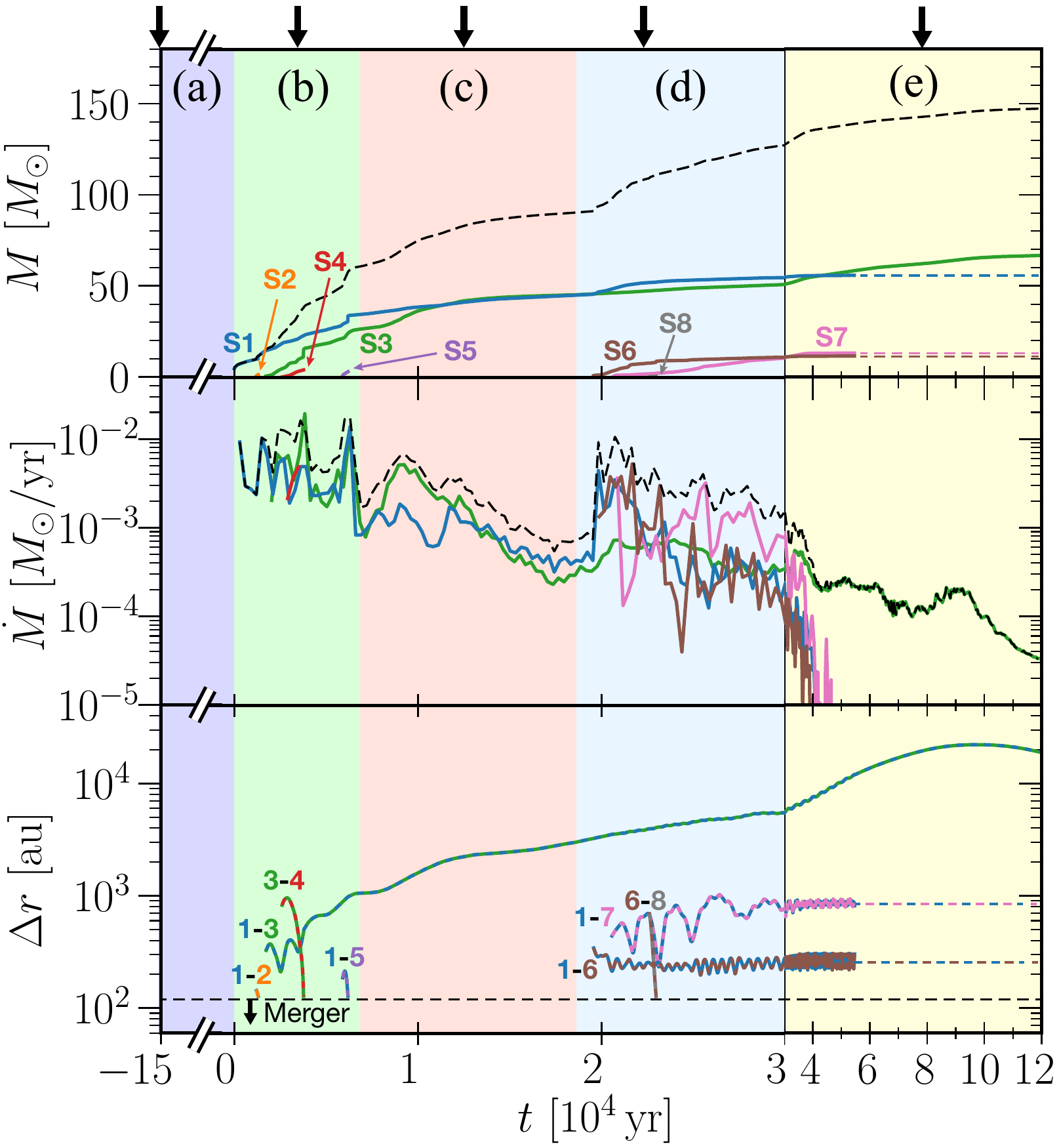} \caption{Time evolution of
the mass (top), accretion rate (middle), and separation (bottom) of each
protostar.  The time $t$ is measured from the first protostar formation,
which happens $1.5\times10^{5}\,\mathrm{yr}$ after the start of the
simulation.  The labels (a)--(e) shown at the top correspond to the
evolutionary phases discussed in the text.  The line colors represent
the IDs of the protostars indicated in the top panel.  In the top and
middle panels, the dashed lines show the total mass and accretion rate,
respectively.  In the bottom panel, we plot the separations of the pairs
of protostars whose IDs are indicated with the pairs of numbers, with
the horizontal dashed line showing the threshold separation for merger.
We do not solve the individual dynamics of S1, S6, and S7 after $t=
5.5\times10^4\,\mathrm{yr}$, as described in the text. The top black
arrows mark the times at which we present snapshots in
Fig.~\ref{fig:pop3_evolution}.  }  \label{fig:sp_evolution}
\end{figure}

\begin{figure}
 \centering \includegraphics[width=8cm]{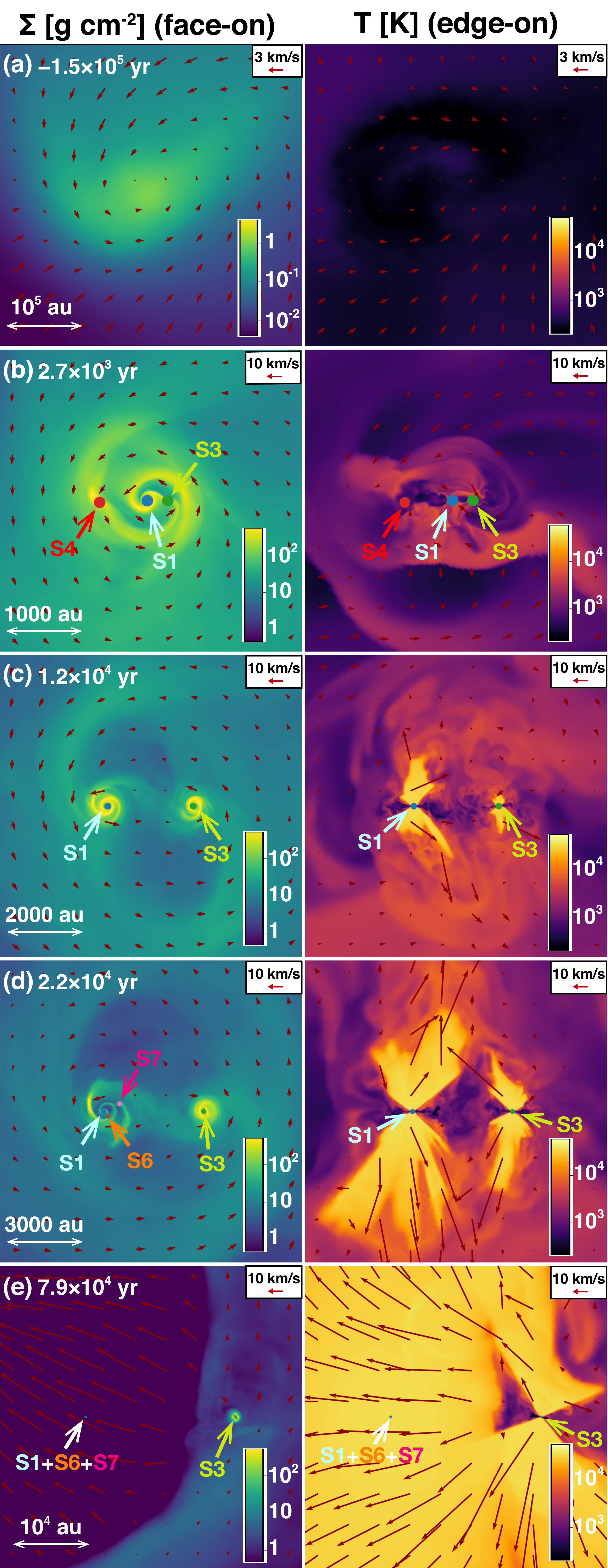}
 \caption{Snapshots of the gas distribution and protostar configuration
 in each of the five evolutionary phases. We show the face-on view of
 the surface density field with density-weighted velocity (left) and the
 edge-on sliced temperature field with velocity (right), along with the
 positions of protostars (thick arrows). The times of snapshots
 presented here are marked with the arrows in
 Fig.~\ref{fig:sp_evolution}. } \label{fig:pop3_evolution}
\end{figure}

\begin{figure}
 \centering \includegraphics[width=7.8cm]{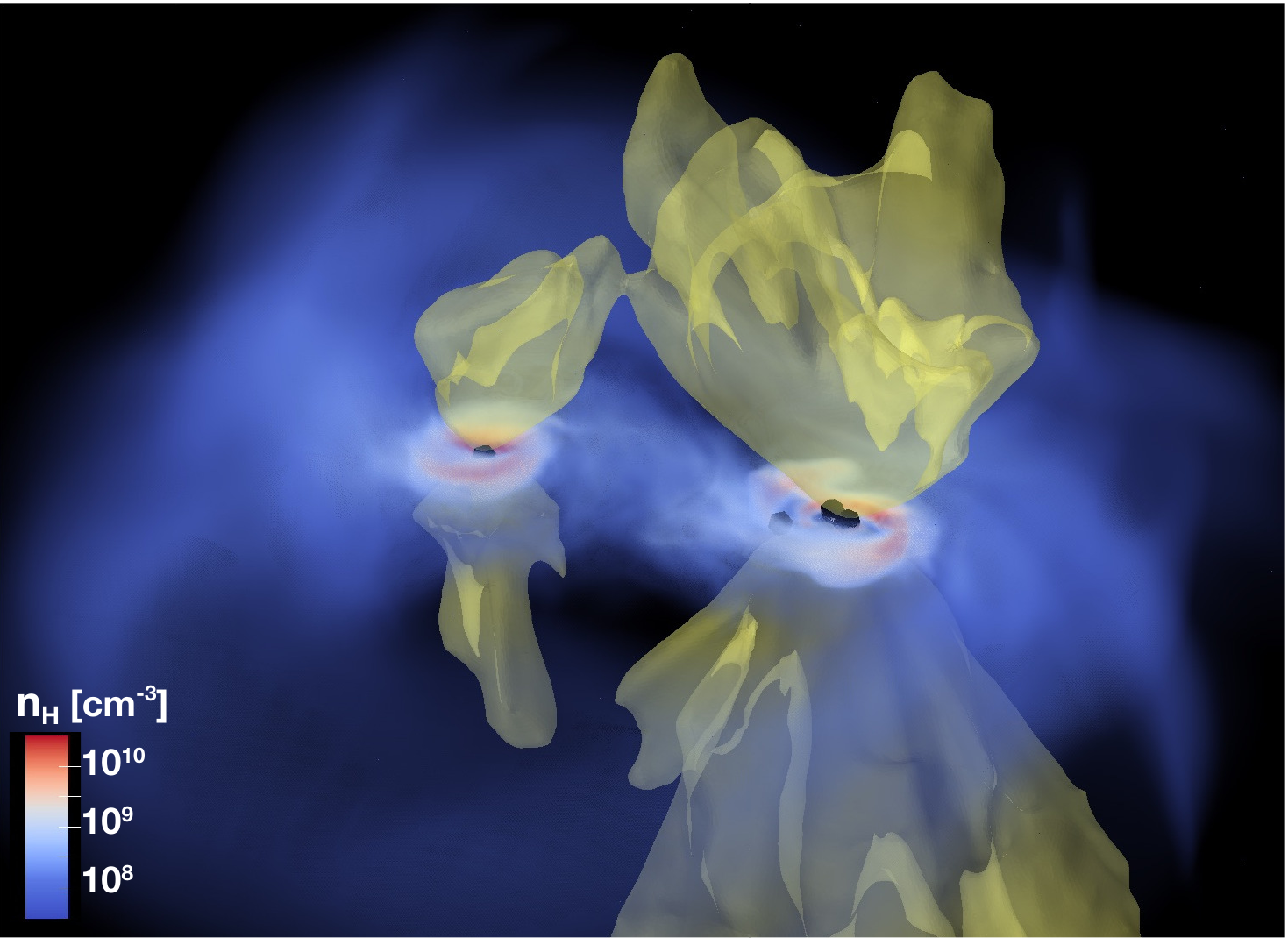}
 \caption{Volume rendering of the density field, together with
 ionization fronts (yellow surfaces) and protostars (black dots),
 $2.2\times10^4\,\mathrm{yr}$ after the first protostar formation (the
 same time as panel (d) of Fig.~\ref{fig:pop3_evolution}).  Bipolar
 ionized bubbles are formed around both of the single-star (left) and
 mini-triplet (right) systems. } \label{fig:3d_rendering}
\end{figure}

Fig.~\ref{fig:sp_evolution} shows the time evolution of the sink
particles, i.e., protostars, appearing in our simulation. In the figure,
masses, accretion rates, and separations are plotted from top to
bottom. We give IDs S1--7 to the protostars in order of the formation
time.  According to qualitative transitions of the system, we define
five evolutionary phases: (a) gravitational collapse, (b) initial
fragmentation, (c) binary accretion, (d) late-time fragmentation, and
(e) photo-evaporation.  For each evolutionary phase, we show a snapshot
of the face-on surface density and edge-on temperature in
Fig.~\ref{fig:pop3_evolution}.  In the late-time fragmentation phase, we
also present a 3D rendering view in Fig.~\ref{fig:3d_rendering}.  Below,
we will describe how star formation proceeds in our simulation, tracing
the five evolutionary stages.
\\[-8pt]

{\bf (a) Gravitational collapse.} We start our simulation from the
initial condition of the cloud $1.5\times10^{5}\,\mathrm{yr}$ before the
first protostar formation (Fig.~\ref{fig:pop3_evolution}a).  The
gravitational collapse proceeds in a self-similar fashion
\citep{Larson:1969aa,Penston:1969aa}, with the central core having an
oblate shape due to the rotation of gas.
\\[-8pt]

{\bf (b) Initial fragmentation.} As the maximum density increases as a
result of the self-similar collapse, the first protostar, S1, forms at
the center of the rotating core.  Subsequently, the gas with finite
angular momentum falls to the vicinity of S1 and a circum-stellar disk
is formed.  The disk is highly gravitationally unstable because of its
relatively high mass compared with the newly forming central protostar
and vigorously fragments into several protostars
(Fig.~\ref{fig:pop3_evolution}b).  Most of them, however, do not survive
as a result of accretion on the central star after inward migration
through the disk or merger with others by three-body scattering.
\\[-8pt]

{\bf (c) Binary accretion.} After the initial fragmentation phase, only
two stars, S1 and S3, survive and make up a binary system.  They are
surrounded by a circum-binary disk and each of them has its own
circum-stellar disk (Fig.~\ref{fig:pop3_evolution}c). The gas accretes
from the circum-binary disk to the circum-stellar disks and each star
acquires the mass from its own circum-stellar disk.  The accretion
drives the binary evolution in mass and orbit: the total mass increases
keeping the mass ratio around unity, and the separation increases due to
accretion of higher angular momentum gas from the circum-binary disk.
At this moment, double bipolar ionized bubbles with $T\gtrsim
3\times10^4\,\mathrm{K}$ grow around the binary stars as their EUV
emissivities rise (Fig.~\ref{fig:pop3_evolution}c, right). Note that
this phenomenon could not be captured by previous simulations, which
could treat only the central radiation source (e.g.,
\citealp{Stacy:2016aa}; H16), and become tractable for the first time
with our multi-source RT with the ART method.  The accretion rate begins
to decrease as the flows in the polar directions are quenched by the
bubbles and the gas supply continues only in the equatorial directions.
\\[-8pt]

{\bf (d) Late-time fragmentation.} Due to imbalance between accretion
rate from the circum-binary disk to the circum-stellar disks and that
from circum-stellar disks to the stars, the gas accumulates on the
circum-stellar disks.  As a result, the circum-stellar disk of S1
fragments into a few objects (Fig.~\ref{fig:pop3_evolution}d and
\ref{fig:3d_rendering}).  After a series of scatter and merger events, a
mini-triplet system emerges, with newly formed companion stars (S6 and
S7) orbiting around the massive central star (S1).  Rapid accretion onto
the companions embedded in the disk, together with the enhanced
accretion onto the central object by non-axisymmetries, quickly exhausts
the disk material.  Due to the hierarchical structure of the
mini-triplet system both in mass and distance, the orbits of the
component stars remain nearly stable.  The ionized bubbles continue
expanding with more ionizing radiation being emitted as the stars become
more massive and with the density in the surrounding envelope decreasing
as the collapse of natal cloud proceeds.
\\[-8pt]

{\bf (e) Photo-evaporation.} Thereafter, the disk in the mini-triplet
system, which has already depleted significantly by the accretion onto
the stars, is lost by the photo-evaporation due to the radiation from
the central massive star.  As a consequence, the ionized bubble around
the mini-triplet expands and merges with that around the star S3
(Fig.~\ref{fig:pop3_evolution}e). The accretion rate onto S3 is
diminished gradually by the radiation not only by S3 itself but also by
the mini-triplet.  The distance between S3 and the mini-triplet becomes
larger partly because the gravitational binding becomes weaker due to
the loss of the gas in between them by the photo-evaporation.  At
$5.5\times10^{4}\,\mathrm{yr}$ after the first protostar formation, we
replace the mini-triplet system with a single sink particle that
represents the gravity center, to save the computational cost.  Since
the properties of the mini-triplet system, namely, the masses and
separations among member stars, hardly changed for the last
$10^{4}\,\mathrm{yr}$, we assume that they remain unchanged in the rest
of the simulation. \\[-8pt]

We stop our simulation at $1.2\times10^5\,\mathrm{yr}$ after the first
protostar formation, when the mass of S3 almost reaches its final value
with the accretion rate already reduced to $\dot{M}\sim 3\times
10^{-5}\,M_\odot\,\mathrm{yr^{-1}}$.  Extrapolating this decreasing
trend of $\dot{M}$ in time, we find the mass of S3 increases by at most
$\lesssim M_{\odot}$.  We thus expect that further mass growth of S3 is
insignificant.  This means that the star formation process has been
practically completed at the end of the simulation.

The end product at the final time step is a massive binary system
consisting of stars with $56$ and $67\,M_\odot$ orbiting each other at a
wide separation of $2\times10^4\,\mathrm{au}$.  The $56\,M_\odot$ star
is also a member of a mini-triplet system with smaller mass companions
with $12$ and $13\,M_\odot$ orbiting at $200$ and $800\,\mathrm{au}$,
respectively.  We have observed the formation of massive Pop III star
binaries starting from the cosmological initial condition.

\section{Discussion} 
\label{sec:discussion}

In this Letter, we have investigated the formation of a Pop III stellar
system with an initial condition of a typical primordial star-forming
cloud, by way of radiation hydrodynamics simulation.  When the radiation
feedback from the forming protostars quenches the gas supply, a binary
system consisting of nearly equal-mass massive stars emerges, with one
of the stars making up its own mini-triplet system with less-massive
close companion stars. Although we have examined only one case in this
Letter, our results suggest that Pop III stars are commonly formed as
massive binary/multiple systems.

In spite of the same initial condition, we have observed the formation
of multiple stars while only a single star was formed in the previous
simulation of H16.  This difference may come from a difference in
resolution. In H16, only a single star survives possibly because of
artificial mergers due to the insufficient resolution around companion
stars. Besides, the total mass in stars ($\sim150\,M_\odot$) in our
simulation is smaller than that of the single star ($\sim300\,M_\odot$)
in H16, which may be related to the fact that the single star always
resides at the center of the cloud, where the accretion is not easily
quenched by the protostellar feedback because of high density.  The
total mass being shared with several stars, the mass of an individual
object is even smaller in our simulation.  This result should be taken
into account in discussing the subsequent evolution of the Universe as
the intergalactic-scale feedback from the Pop III stars, e.g., types of
SNe \citep[e.g.,][]{Woosley:2002aa}, largely depends on their masses.

The number of stars obtained in our simulation should be regarded as the
lower limit because of our usage of the sink particle technique and
ignorance of what may happen inside the sink particles, such as
fragmentation.  Moreover, although we assume that overlapping sink
particles merge, two or more stars may continue orbiting each other at a
distance shorter than the sink radius.  Each sink particle, however, is
likely to contain at least one massive star because a system consisting
only of a large number of low-mass stars is unstable due to mutual
gravitational scatterings.  Therefore, our conclusion that Pop III stars
typically form as massive binary/multiple systems is not altered by
the possible existence of unresolved objects.

The binary system with $60$ and $70\,M_\odot$ stars found in our
simulation is massive enough to be a progenitor of the binary BH mergers
observed in gravitational waves
\citep[e.g.,][]{Kinugawa:2014aa,Hartwig:2016aa,Abbott:2016aa,Belczynski:2017aa}
while the separation is too wide for any interaction to shrink the orbit
in the late stage of stellar evolution and the remnants are unlikely to
merge within the age of the Universe.  To examine whether Pop III stars
can be the progenitors of the observed BH binaries, the distribution of
binary separations needs to be known.  This is also crucial for
predicting the abundance of Pop III X-ray binaries, which contribute to
heating up the intergalactic medium in the epoch of reionization.  For
this purpose, we plan to perform similar simulations with a large number
of samples in the future.  Note that both the masses and separation of
our binary system are located near the upper end of the statistical
distributions of Pop III binaries given by \cite{Stacy:2013aa}, who took
their samples $5\times10^{3}\,\mathrm{yr}$ after the first protostar
formation.  The masses and separation of our system are already on the
large side in the distributions by \cite{Stacy:2013aa} at the same
timing and become even larger during the later evolution (see
Fig.~\ref{fig:sp_evolution}).

Aside from the massive wide binary, we have also seen the formation of
the mini-triplet system.  Its separation is much smaller than that of
the massive binary because it is formed from the circum-stellar disk,
which has lower angular momentum compared with the original cloud.  On
the contrary, the massive binary has a large separation as a result of
the accretion of high angular momentum gas from the cloud.  In fact, we
have seen that the separation of the binary increases as it accretes the
gas from the circum-binary disk, as suggested in recent simulations of
binary accretion (\citealp{Munoz:2019aa,Moody:2019aa,Duffell:2019aa}).
The late-time disk fragmentation leading to such multiple systems may
play some roles in formation of close binaries, which can evolve to the
progenitors of BH merger events or X-ray binaries.  Additionally, other
mechanisms, such as a-few-body scatterings of protostars and angular
momentum extraction by magnetic fields, if any, might help shape the
close binaries.

We have succeeded in seeing a new evolutionary aspect of the Pop III
star formation.  This is, however, just the beginning of the attempts
toward its thorough understandings as some more processes, such as
sub-sink scale physics, are still to be clarified.  Those issues need to
be addressed in future studies.\\


The authors thank Sunmyon Chon, Michiko Fujii, Masahiro Machida, Massimo
Ricotti, Hajime Susa, Hidekazu Tanaka, Ataru Tanikawa, Masauyuki
Umemura, Hidenobu Yajima, and Naoki Yoshida for fruitful discussions and
comments.  K.S. and S.H. appreciate the support by the JSPS Overseas
Research Fellowship and JSPS Research Fellowship, respectively.  This
work is supported in part by MEXT/JSPS KAKENHI Grant Number 17H02863,
17K05394, 18H05436, 18H05437 (T.M.), 16H05996, 19H09134 (T.H.), 18J01296
(S.H.), and 17H01102, 17H02869, 17H06360 (K.O.).  The numerical simulations
were performed on the Cray XC50 at CfCA of the National Astronomical
Observatory of Japan, the computer cluster {\tt Draco} at Frontier
Research Institute for Interdisciplinary Sciences of Tohoku University,
and the Cray XC40 at Yukawa Institute for Theoretical Physics in Kyoto
University.

\bibliographystyle{apj}



\end{document}